\documentclass[aps,prl,twocolumn,amssymb,showpacs]{revtex4}
\usepackage{epsfig}

\newcommand{\Maybeeq}[1]{Eq.~(#1)}
\newcommand{\Maybeeqs}[1]{Eqs.~(#1)}
\newcommand{\maybeeq}[1]{Eq.~(#1)}
\newcommand{\maybeeqs}[1]{Eqs.~(#1)}
\newcommand{\Figs}{Figs.}
\newcommand{\figs}{Figs.}

\newcommand{\fig}{Fig.}

\newcommand{\cc}{{\rm c}}
\newcommand{\op}{{\rm op}}
\newcommand{\lcftbdy}{\partial\theta\partial\bar\theta}
\newcommand{\lcftbulk}{\partial\theta\bar\partial\bar\theta+
\bar\partial\theta\partial\bar\theta}

\begin{document}

\title{The four height variables
of the Abelian sandpile model}

\author{M. Jeng}

\email{mjeng@siue.edu}

\affiliation{
Box 1654, Department of Physics, Southern Illinois
University Edwardsville, Edwardsville, IL, 62025}

%-----------------------------------------------------------------%

\begin{abstract}

\noindent We study the four height
variables in the Abelian sandpile model. We argue that
the four variables are not represented by the
same operator along closed boundaries, or in the bulk.
Along open boundaries, we calculate all n-point
correlations, and find that there, all height
variables are represented by the same operator. We introduce
dissipative defect points, and show that along open
boundaries they are represented by the same operator
as the height variables. We show that dissipative defect
points along closed boundaries, or in the bulk, have no
effect on weakly allowed cluster variables.

\end{abstract}


\maketitle

%-----------------------------------------------------------------%

The Abelian sandpile model (ASM) proposed by Bak, Tang and
Wiesenfeld produces power laws without any 
fine-tuning of parameters, and thus potentially provides
an explanation for how power laws can arise in
nature~\cite{BTW}. Since its introduction, the ASM has been
used to analyze a diverse range of
systems---see \cite{BakBook} for a review.

The ASM is an extraordinarily simple mathematical
model (see~\cite{BTW} for a description).
However, significant aspects of the ASM
are still unknown.
While the height one variable is
well understood, the
higher height variables (two, three and four) are not.
We approach this problem in this paper
primarily by looking at correlation functions along or 
near open and closed boundaries.
We also investigate the role of dissipative 
defect sites.

The correlation functions are computed with an elegant
method, introduced by Majumdar and
Dhar, who used it to calculate unit height
probabilities and correlations~\cite{Dhar.UnitCorrelations}. 
The recurrent states of the ASM can be mapped to
spanning trees drawn on the 
sandpile lattice~\cite{DharFirst,Dhar.CFT}.
A spanning tree is a connected, directed, acyclic
graph, such that every site has a path leading to the
root, which 
is ``off the edge of the sandpile'' (i.e. the ``site'' connected 
to all open boundaries).  The site $i$ is said to be a
predecessor of the site $j$ if the path from $i$ to the root
goes through $j$. 
The probability for a site $i$ to have height $h$ is
equal to the probability that in a spanning tree,
exactly $h-1$ of its nearest neighbors will be
predecessors of $i$ (NNP's, for nearest-neighbor
predecessors)~\cite{Priezzhev}.

Certain height configrations in the ASM correspond to local
restrictions on the spanning tree. For example, the condition 
for a site to have height one is equivalent to a spanning 
tree condition that it be disconnected from three of its 
neighbors.  
Any probabilities corresponding to
{\it local} restrictions on the spanning tree
(known as weakly allowed cluster variables) can be 
calculated as finite-dimensional matrix determinants
with the Majumdar-Dhar 
method~\cite{Dhar.UnitCorrelations,DharFirst,Dhar.CFT}.
These calculations require use of
the lattice Green function.

Spanning trees are described by the central charge $-2$
logarithmic conformal field theory ($c=-2$ LCFT), and so 
the question arises as to how variables in the ASM should 
be identified with operators in the LCFT.  Mahieu and Ruelle, 
using the Majumdar-Dhar methods, calculated correlation 
functions for a number of height configurations, and showed
that they agree with LCFT correlations, with appropriate 
field identifications~\cite{Mahieu.Ruelle}. However, these 
methods do not allow the calculation of probabilities or
correlations for bare higher height variables,
because they are associated with predecessor 
relationships, which are nonlocal; a site $i$ can be a predecessor 
of a neighbor $j$ by a long path which goes far from 
either $i$ 
or $j$.  
Priezzhev has 
developed $\theta$-graph methods to find the bulk probabilities 
for higher height variables~\cite{Priezzhev}. 
However, the bulk
correlations for higher height variables remain unknown.

The sandpile boundaries can either be open, where grains
of sand can fall off the edge during topplings, or closed, 
where they cannot. Ivashkevich has shown that along either
boundary type, the nonlocal arrow diagrams
associated with higher height probabilities and two-point 
correlations can be written as linear combinations 
of local arrow diagrams~\cite{Ivashkevich}. He found that 
all boundary two-point correlation functions, between all height 
variables, fall off as $1/r^4$, and thus 
argued that all height variables correspond to the same 
LCFT field. We reanalyze his calculations and results below.
Mahieu and Ruelle presented other evidence 
that the higher height variables are identical to the
unit height variable in the scaling limit; however, they
also pointed out that this identification appears
inconsistent with LCFT operator product 
expansions (OPEs)~\cite{Mahieu.Ruelle}. We argue here that 
analysis of boundary correlation functions, and
heights far from the boundary,
indicate that the
height variables should {\it not} all receive the same field
identification along closed boundaries, or in the bulk. 
However, along open
boundaries, they are identical in all n-point
correlations; additionally, dissipative defect
points
along open boundaries receive the same field
identification as the height variables.

\noindent {\bf Closed:} We define, for all $n$-point
correlation functions along closed boundaries,

\begin{eqnarray}
\nonumber
f_\cc (a_1,a_2,\dots,a_n) & = & \\
& & 
\hspace{-1.0 in} 
<(\delta_{h_{x_1},a_1}-p_{a_1,\cc})
\dots
(\delta_{h_{x_n},a_n}-p_{a_n,\cc})>_\cc \ ,
\end{eqnarray}

\noindent where ``$\cc$'' stands for ``closed,''
the $a_i$'s are heights, $h_x$ is the height at position $x$
along the boundary, and
$p_{a_i,\cc}$ is the constant probability for a site on the
closed boundary to have
height $a_i$. 
Note that along closed boundaries, sites
cannot have height four. Ivashkevich
found 

\begin{eqnarray}
\label{eq:twoPoint.11}
f_\cc (1,1) & = & \left(-{9 \over \pi^2} + {48 \over \pi^3} 
- {64 \over \pi^4} \right) {1 \over {(x_1-x_2)^4}} +\dots \\
\label{eq:twoPoint.22}
f_\cc (2,2) & = & \left(-{61 \over {4 \pi^2}} + {96 \over \pi^3} 
- {144 \over \pi^4} \right) {1 \over {(x_1-x_2)^4}} +\dots \\
\label{eq:twoPoint.33}
f_\cc (3,3) & = & \left(-{1 \over {4 \pi^2}} + {8 \over \pi^3} 
- {16 \over \pi^4} \right) {1 \over {(x_1-x_2)^4}} +\dots
\end{eqnarray}

\noindent While these correlations all have the same
power law, the coefficient in $f_\cc (1,1)$ is
negative, while the coefficients of $f_\cc (2,2)$ 
and $f_\cc (3,3)$ are positive.
These sign differences indicate that the three height variables are
in fact not all represented by the same field operator.
Furthermore, the coefficients of $f_c(a,a')$
in~\cite{Ivashkevich} do not factorize (into $K_a K_{a'}$),
as would be expected if all three height variables were
the same up to rescaling.

As a check, we have rederived all results 
of~\cite{Ivashkevich}.
(There is a misprint in the result for
$f_\cc (3,3)$ in~\cite{Ivashkevich}.)
While we agree with the results
of~\cite{Ivashkevich}, the analysis there
appears to have several errors. For the two-point
correlations between $r_i$ and $r_j$, it divides ASM states
into sets $S_{ab}$, where the states of $S_{ab}$
are allowed with heights
$h_{r_i}\geq a$ and $h_{r_j}\geq b$,
but forbidden otherwise.
However, not all ASM states fall into such sets. For
example,
there are states allowed when $(h_{r_i},h_{r_j})=(1,2)$,
and allowed when $(h_{r_i},h_{r_j})=(2,1)$, yet
forbidden when $(h_{r_i},h_{r_j})=(1,1)$.
However, the relationship in~\cite{Ivashkevich}
between the $S_{ab}$
and the spanning trees
is also not quite correct, and
these errors end up largely, but not entirely,
cancelling.
The spanning tree representation is somewhat 
surprising.
The natural assumption (made in~\cite{Ivashkevich})
is that
all spanning trees where $r_i$ and $r_j$ each have one NNP
contribute to the 2-2 correlation in the ASM. However there
are some such spanning trees that do not; for example, trees
where $r_i$ and $r_j$ each have one NNP, and neither
is a predecessor of the other, but 
a neighbor of $r_i$ is a predecessor of $r_j$, and a
neighbor of $r_j$ is a predecessor of $r_i$, contribute
not to the 2-2 correlation, but to the 2-3 correlation.
Such cases 
can be written as linear combinations of closed loop
diagrams, and then calculated using a generalized
Kirchoff theorem~\cite{Priezzhev}. Luckily, these graphs
fall off as $1/(x_1-x_2)^6$, 
leaving the
results of~\cite{Ivashkevich} unaffected.

We have calculated all closed boundary
three-point correlation functions
that include at least one unit
height variable. Some results are

\begin{eqnarray}
\label{eq:f111}
f_\cc (1,1,1) & = & {{2(3\pi-8)^3} \over
{\pi^6 (x_1-x_2)^2(x_1-x_3)^2(x_2-x_3)^2}}
+\dots \\
\nonumber
f_\cc (1,1,2) & = & - {{8(\pi-3)(3\pi-8)^2} \over
{\pi^6 (x_1-x_2)^2(x_1-x_3)^2(x_2-x_3)^2}} \\
& & - {{(3\pi-8)^2} \over {\pi^5
(x_1-x_3)^3(x_2-x_3)^3}}+\dots \\
\nonumber
\label{eq:f122}
f_\cc (1,2,2) & = & - {{4(3\pi-8)(-5\pi^2+39\pi-72)} \over
{\pi^6 (x_1-x_2)^2(x_1-x_3)^2(x_2-x_3)^2}} \\
& & + {{(3\pi-8)(24-7\pi)} \over {2\pi^5 (x_1-x_2)^3(x_1-x_3)^3}}
+\dots
\end{eqnarray}

\noindent The correlation functions must satisfy identities
resulting from the fact that the height probabilities at any
site must sum to one. For example,
$\sum_{a_3=1}^3 f_\cc (a_1,a_2,a_3)=0$. As a check, we have 
verified that all other three-point correlations that
include at least one unit height variable satisfy all such
identities. All calculated two- and three-point
correlations are consistent with bulk correlations
identifying
$-({2(3\pi-8)}/ {\pi^2})
\left( \partial\theta\partial\bar\theta \right)$
with the height one variable, 
$({6(\pi-4)}/{\pi^2})
\left( \partial\theta\partial\bar\theta \right)
+(1/2\pi) \theta\partial^2\bar\theta$
with the height two variable, and 
$(8/{\pi^2})
\left( \partial\theta\partial\bar\theta \right)
-(1/2\pi) \theta\partial^2\bar\theta$
with the height three variable.
There are no $\bar\partial$'s, since
fields near a boundary of a (L)CFT can be thought of
as purely holomorphic; CFT boundary correlations
behave like bulk correlations, where the 
antiholomorphic parts of the
field act like holomorphic pieces at mirror locations
across the 
boundary~\cite{Cardy.mirror,LCFT.general,LCFT.general.2}.
It is also consistent to make the substitution
$\theta\rightarrow\bar\theta$, 
$\bar\theta\rightarrow -\theta$ in all field
identifications above, as the $c=-2$ CFT is symmetric
under this transformation.
($\theta$ and
$\bar\theta$ are variables in the $c=-2$ LCFT.
See~\cite{Mahieu.Ruelle}
for a brief description of the $c=-2$ LCFT, and references
therein for a more complete treatment.)

\begin{figure}[tb]
\epsfig{figure=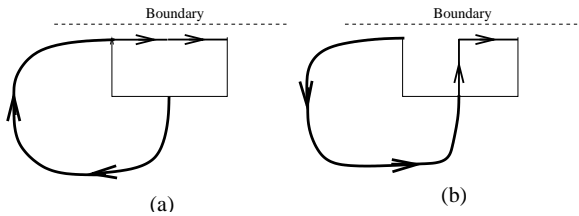,width=3.0in}
\caption{Two graphs identical only in one-point functions}
\label{fig:phi1andphi2}
\end{figure}

\begin{figure}[tb]
\epsfig{figure=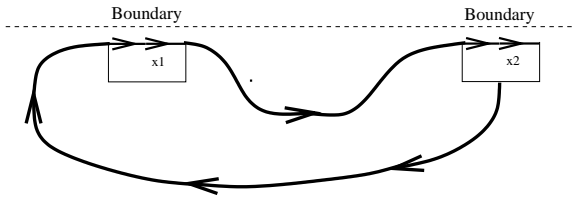,width=3.0in}
\caption{A long path}
\label{fig:looppath}
\end{figure}

We have not been able to calculate all terms
contributing to $f_\cc (2,2,2)$, but note that if
complications arising from graphs such as those
(to be) discussed below 
\figs~\ref{fig:phi1andphi2} and~\ref{fig:looppath}
are ignored, we obtain

\begin{equation}
f_\cc (2,2,2) = -  {{(24-5\pi)(-576+384\pi-61\pi^2)} \over
{4\pi^6(x_1-x_2)^2(x_1-x_3)^2(x_2-x_3)^2}} +\dots
\end{equation}

\noindent which is consistent with the field
identification above
for the height two variable above.

We have also introduced a dissipative defect
site on the closed boundary. At a normal site of a closed boundary,  
a site topples whenever it has more than three grains, losing 
three grains, and sending one grain to each of its three 
neighbors. On an open boundary, 
sites topple when they have more than four grains, and thus
send an extra grain
off the edge.  We add a ``dissipative
defect point'' on the closed boundary, 
where the site can
now have up to $3+k$ grains, and where $k$ grains 
fall off the edge with each toppling.  
Previously, dissipation has been added uniformly in the bulk,
where it breaks criticality~\cite{dissip.nocrit}.
(But see~\cite{Mine.BC.Univ} for dissipation along a line.)
We have shown that if the defect-free 
closed boundary Green function is $G_{\cc,0}(\vec{r_1},\vec{r_2})$, 
then a defect at $\vec{d}$ changes the Green function
(dropping constant terms) to

\begin{equation}
G_\cc (\vec{r_1},\vec{r_2}) =
G_{\cc,0}(\vec{r_1},\vec{r_2}) - 
G_{\cc,0}(\vec{r_1},\vec{d}) - 
G_{\cc,0}(\vec{d},\vec{r_2})
\label{eq:Green.defect.open}
\end{equation}

\noindent The modified Green function is
independent of the value of $k$.
(This is reasonable, since regardless of the amount of 
dissipation, the defect provides the only possible
spanning tree route to the root.)
Using this new Green function, for a defect at $x=0$ on
the boundary, we find 

\begin{eqnarray}
\label{eq:defectnochange}
f_\cc (1) & = & 0 \\
\label{eq:defect2}
f_\cc (2) & = & - {1 \over {2\pi x_1^2}} + \dots 
\end{eqnarray}

\noindent The height one probability 
is unaffected, {\it to all orders}, at all points,
in the bulk and the boundary, by the 
defect
(except at the
defect itself).
$\sum_{a=1}^3f_c(a)=0$, so $f_c(3)=-f_c(2)$.
\Maybeeqs{\ref{eq:defectnochange}-\ref{eq:defect2}} 
have been numerically confirmed.

These results indicate that a dissipative defect on a closed
boundary is represented by a dimension 0 operator.  
(Note that Mahieu and Ruelle have identified uniform dissipation in
the bulk of the ASM with the integral of
a dimension 0 operator~\cite{Mahieu.Ruelle}.)
The defect has correlations with the height two and three 
operators, but not with the height one operator, again 
pointing to different field identifications for the different 
height variables. Since they differ along closed
boundaries, they must also differ in the bulk, since
CFT boundary operators are derived from OPE's of bulk
operators~\cite{Cardy.mirror}. 

Finally, we have found the correlation function of $n$ 
unit height 
variables, at positions $x_i$ ($i=1,2,\dots,n$) along 
the boundary. This requires local arrow constraints at $3n$ 
vertices of the ASM, and thus the calculation of a
$3n$-dimensional matrix determinant. The matrix is divided 
into 3 by 3 block submatrices, such that the diagonal blocks 
are all identical, and the off-diagonal blocks all have 
the same form. A rotation makes the matrix 
diagonal in 2 out of every 3 rows (and columns).  The universal 
part of the correlation function is 
found to be 

\begin{equation}
\label{eq:unit.closed.corr}
\left({{3\pi-8}\over\pi^2}\right)^n 
{\rm det} \left( {\bf M} \right) \ ,
\end{equation}

\noindent where {\bf M} is the $n$-dimensional matrix

\begin{equation}
\label{eq:MMatrix}
M_{ij} = \left\{ 
\begin{array}{cl}
0 & \rm{if}\ i=j \\
1/(x_i-x_j)^2 & \rm{if}\ i\neq j
\end{array}
\right.
\end{equation}

This is the same as the $n$-point correlation of
$-({2(3\pi-8)}/ {\pi^2})
\left( \partial\theta\partial\bar\theta \right)$
in the bulk,
confirming the unit height identification 
below \maybeeq{\ref{eq:f122}}.
In~\cite{Mahieu.Ruelle}, the unit height variable 
in the bulk was
associated with $\lcftbulk$. 

\noindent {\bf Open:} The correlation functions are simpler
along open boundaries. We define the operators

\begin{equation}
\phi_a(x) \equiv {{\delta_{h_x,a}-p_{a,{\op}}} \over K_a} \qquad,
\rm{where}\ 
a=1,\dots, 4
\end{equation}

\noindent $p_{a,{\op}}$ is the constant probability for a site along 
an open boundary to have height $a$ (already found 
in~\cite{Ivashkevich}), and the $K_a$ are normalization
factors:

\begin{eqnarray}
\begin{array}{ll}
p_{1,{\op}}={9\over 2}-{42\over\pi}+
{320\over{3\pi^2}}-{512\over{9\pi^3}} &
K_1=-{3\over\pi}+{80\over{3\pi^2}}-{512\over{9\pi^3}} \\ \\
p_{2,{\op}}=-{33\over 4}+{66\over\pi}- 
{160\over\pi^2}+{1024\over{9\pi^3}} &
K_2={9\over\pi}-{200\over{3\pi^2}}+{1024\over{9\pi^3}} \\ \\
p_{3,{\op}}={15\over 4}-{22\over\pi}+
{160\over{3\pi^2}}-{512\over{9\pi^3}} & 
K_3=-{7\over\pi}+{40\over\pi^2}-{512\over{9\pi^3}} \\ \\
p_{4,{\op}}=1-{2\over\pi} & 
K_4={1\over\pi}
\end{array}
\end{eqnarray}

\noindent Using methods similar to those used to find
\maybeeq{\ref{eq:unit.closed.corr}}, we find that the
$n$-point open boundary correlation
function,

\begin{equation}
\label{eq:nptcorr}
\left<\phi_{a_1}(x_1)\phi_{a_2}(x_2)\dots\phi_{a_n}(x_n)\right> 
\,
\end{equation}

\noindent is equal to det({\bf M}).
(These results reproduce and extend the 
one- and two-point functions found 
in~\cite{Ivashkevich}.)
\Maybeeq{\ref{eq:nptcorr}} is
independent of the $a_i$'s, showing that the
four height variables, upon rescaling, do all receive the
same field assignment ($\lcftbdy$)
along open boundaries. Apparently the
different height variables correspond to different fields
in the bulk, and remain different along closed
boundaries, but become the same along open
boundaries. 

We again add dissipative defects. At the site $\vec{d}=(x,y)$, 
we increase the toppling condition by $k>0$, so that $k$ 
grains of sand are dissipated with each toppling.  
We assume that all open defect
sites are $y=\mathcal{O}(1)$ from the boundary (at $y=1$).
Then the new Green function is

\begin{eqnarray}
\nonumber
G_{\op}(\vec{r_1},\vec{r_2}) & = &
G_{\op,0}(\vec{r_1},\vec{r_2}) \\
& & \hspace{-0.52in} - {k\over{1+kG_{\op,0}(\vec{d},\vec{d})}}
G_{\op,0}(\vec{r_1},\vec{d}) G_{\op,0}(\vec{d},\vec{r_2})
\label{eq:Green.defect.closed}
\end{eqnarray}

\noindent 
$G_{\op,0}(\vec{r_1},\vec{r_2})$
is the defect-free open boundary Green function.
\Maybeeqs{\ref{eq:Green.defect.closed}}
and (\ref{eq:Green.defect.open}) are different because 
the Green function between nearby points is $\mathcal{O}(1)$ 
on an open boundary, but $\mathcal{O}(\ln L)$ on a closed 
boundary, where $L$ is the distance to the nearest open 
boundary.  We have generalized
\maybeeq{\ref{eq:Green.defect.closed}} 
for multiple open dissipative defects.

We define $\phi_5(\vec{d};k)$ 
as the operator corresponding to the addition of a
defect of strength $k$ at $\vec{d}=(x,y)$, and then the
multiplication of all correlation functions
containing $\phi_5(\vec{d};k)$ by
$\pi(1+kG_{\op,0}(\vec{d},\vec{d}))/(ky^2)$. 
Then the connected $n$-point correlation function in 
\maybeeq{\ref{eq:nptcorr}} is still given by the
connected terms of det({\bf M}),
even if some of the $a_i$ are now equal to 5, and regardless
of the (different) values of the $k$'s at the various 
defects.  
So the addition of a local
dissipative defect near an open boundary is, like the height
variables, represented by $\lcftbdy$.

The open boundary is much more
tractable than the closed boundary for several
reasons.
For calculating one-point functions on 
any boundary,
when we write the nonlocal arrow diagrams as 
linear combinations of local arrow diagrams, we
use the equivalence of certain nonlocal arrow diagrams.  
For example, Ivashkevich pointed out that 
\Figs~\ref{fig:phi1andphi2}a and~\ref{fig:phi1andphi2}b
are equally likely because we can switch from
\ref{fig:phi1andphi2}a to \ref{fig:phi1andphi2}b
by reversing arrows along the long path~\cite{Ivashkevich}. However, 
\figs~\ref{fig:phi1andphi2}a and~\ref{fig:phi1andphi2}b
are not equivalent if they are embedded in a 
correlation function, which has arrow conditions 
in distant parts of the lattice, because reversing the
long path can change these distant arrows.
This produces extra terms, such as the graph in 
\fig~\ref{fig:looppath}. 

The graph in \fig~\ref{fig:looppath} can be written as 
a linear combination of closed loop graphs.
It, and all analogous
graphs, turn out to fall off faster 
than $1/(x_1-x_2)^4$ for the open case, and thus make no 
contribution to any universal correlation
functions. 
On the other hand, along 
closed boundaries, the graph in \fig~\ref{fig:looppath},
and other analogous graphs, fall off as 
$1/(x_1-x_2)^4$, 
complicating matters (although, in the end, the 
$1/(x_1-x_2)^4$ parts of these closed loop paths end
up cancelling).

Also, the Green function
decays as $1/x^2$ along open boundaries
(unlike the closed and bulk cases, where it grows as
$\ln (x)$~\cite{Bdy.Falloff}).
This allows trace formulae found 
by Mahieu and Ruelle for 2- and 3-point correlations
~\cite{Mahieu.Ruelle}
to be generalized for open boundary $n$-point functions,
simplifying matters.

\noindent {\bf Bulk:} Analysis of the higher height
probabilities well in the bulk, at distances $y\gg 1$ from a
boundary provides a further argument
that the higher height variables should not all receive the
same bulk field identification. Suppose that they
{\it do} all have the same field
identification, up to rescaling. Then we would know both
that

\begin{equation}
p_a (y) = p_{a,{\rm B}}+{c_a \over y^2}+\dots
\end{equation}

\noindent for all $a=1\dots 4$ (based on results for $a=1$
in~\cite{Bdy.Falloff}), and also that all
height $a$-height $a$ correlations in the bulk would have
to have the same sign (negative~\cite{Dhar.UnitCorrelations}).
Upon rescaling the height variables to give them
(negative) unit norm, the $c_a$ are rescaled to
$\tilde{c}_a$. Based on general (L)CFT arguments, the
$\tilde{c}_a$ should be universal numbers, 
independent of 
$a$~\cite{LCFT.general,LCFT.general.2,Cardy.UnivCoeff}. 
(See~\cite{Mine.BC.Univ} for an explicit demonstration of
this, in the context of the ASM.)
Then the original (unrescaled) $c_a$ must all have
the same sign.
However, the four
$c_a$'s cannot all have the same sign, since 
$\sum_{a=1}^4 p_a (y)=1$ for all $y$.
So, contrary to our assumption,
the four height variables must have different
bulk field identifications.  

A dissipative defect in the bulk modifies the Green function
exactly as in \maybeeq{\ref{eq:Green.defect.open}}, and
again does not modify unit height probabilities at
any sites other than the defect. 
However,
modest numerical simulations show that 
higher height probabilites are changed near the
defect. This parallels the the closed case, 
again indicating different field identifications for the 
higher height variables.

More generally, we can show that no height configuration
probabilities that can be calculated by the removal of 
a set of bonds 
in the ASM
(i.e. 
the weakly allowed cluster 
variables, and their correlations)
are affected
by a dissipative defect in the bulk or closed cases.
This lack of correlations suggests that 
the analysis of weakly allowed cluster
variables in~\cite{Mahieu.Ruelle},
while impressive, should
not be expected to give a fully
representative picture of the field structure of the ASM. 

\acknowledgments{After
the bulk of this work was completed, we were
informed of independent, then unpublished,
calculations of
\maybeeqs{\ref{eq:f111}-\ref{eq:f122}}, in the
massive case, by G. Piroux and P.
Ruelle~\cite{PR.parallel}.
Discussions with G. Piroux and P. Ruelle
then led us to the field identifications 
below \maybeeq{\ref{eq:f122}},
and they further corrected an error we had made in these
same
equations.
This work was supported by
Southern Illinois University Edwardsville.
We thank V. Gurarie and E. V. Ivashkevich
for useful discussions.}

%-----------------------------------------------------------------%


\begin{thebibliography}{99}

\bibitem{BTW} P. Bak, C. Tang, and K. Wiesenfeld, Phys. Rev.
Lett., {\bf 59} 381 (1987).

\bibitem{BakBook} P. Bak, {\it How Nature Works} (Oxford
Univ. Press, Oxford, 1997).

\bibitem{Dhar.UnitCorrelations} S. N. Majumdar and D. Dhar,
J. Phys. A: Math. Gen. {\bf 24} L357 (1991).

\bibitem{DharFirst} D. Dhar, Phys. Rev. Lett. {\bf 64} 1613
(1990).

\bibitem{Dhar.CFT} S. N. Majumdar and D. Dhar, Physica
A {\bf 185} 129 (1992).

\bibitem{Priezzhev} V. B. Priezzhev, J. Stat. Phys. {\bf 74}
955 (1994).

\bibitem{Mahieu.Ruelle} S. Mahieu and P. Ruelle,
Phys. Rev. E {\bf 64} 066130 (2001).

\bibitem{Ivashkevich} E. V. Ivashkevich, J. Phys. A:
Math. Gen. {\bf 27} 3643 (1994)

\bibitem{Cardy.mirror} J. L. Cardy, Nucl. Phys. B {\bf 240}
514 (1984).

\bibitem{LCFT.general} I. I. Kogan and J. F. Wheater, Phys.
Lett. B {\bf 486} 353 (2000).

\bibitem{LCFT.general.2} S. Moghimi-Araghi and
S. Rouhani, Lett. Math. Phys.
{\bf 53} 49 (2000).

\bibitem{dissip.nocrit} T. Tsuchiya and M. Katori, 
Phys. Rev. E {\bf 61} 1183 (2000).

\bibitem{Mine.BC.Univ} M. Jeng, ``Boundary conditions and
defect lines in the Abelian sandpile model,"
cond-mat/0310605.

\bibitem{Bdy.Falloff} J. G. Brankov, E. V. Ivashkevich, and
V. B. Priezzhev, J. Phys. I France {\bf 3} 1729 (1993).
%1729-1740

\bibitem{Cardy.UnivCoeff} J. L. Cardy and D. C. Lewellen,
Phys. Lett. B {\bf 259} 274.

\bibitem{PR.parallel} G. Piroux and P. Ruelle, ``Boundary
height fields in the Abelian sandpile model,''
hep-th/0409126, to be published in Physica A.

\end{thebibliography}
\end{document}